\documentclass[a4paper]{article}
\usepackage{amscd,amsmath,amssymb}

\newcommand{\mxi}{\boldsymbol{\xi}}
\newcommand{\mbxi}{\bar{\boldsymbol{\xi}}}
\newcommand{\mpsi}{\boldsymbol{\psi}}
\newcommand{\mbpsi}{\bar{\boldsymbol{\psi}}}
\newcommand{\mlambda}{\boldsymbol{\lambda}}
\newcommand{\mblambda}{\boldsymbol{\bar\lambda}}

\newcommand{\bb}[1]{ \boldsymbol{#1}{}}
\newcommand{\tr}{\;\textrm{Tr}\;}

\newcommand{\p}[1]{(\ref{#1})}

\newcommand{\cF}{{\cal F}}

\newcommand{\cH}{{\cal H}}

\newcommand{\cQ}{{\cal Q}}
\newcommand{\cbQ}{\overline{\cal Q}}

\newcommand{\cL}{{\cal L}}

\newcommand{\bD}{{\overline D}{}}

\newcommand{\bQ}{{\overline Q}{}}

\newcommand{\bB}{{\overline B}{}}

\newcommand{\bxi}{{\bar\xi}}

\newcommand{\bpsi}{{\bar\psi}{}}

\newcommand{\blambda}{{\bar\lambda}}

\newcommand{\cN}{{ {\cal N}   }}

\newcommand{\tPi}{{\widetilde \Pi}}

\newcommand{\und}{\qquad\textrm{and}\qquad}
\renewcommand{\=}{\ =\ }

\newcommand{\be}{\begin{equation}}
\newcommand{\ee}{\end{equation}}
\newcommand{\bea}{\begin{eqnarray}}
\newcommand{\eea}{\end{eqnarray}}

\newcommand{\ba}{\begin{array}} \newcommand{\ea}{\end{array}}

\def\im{{\rm i}}
\def\sfrac#1#2{{\textstyle\frac{#1}{#2}}}

\newcommand{\nn}{\nonumber}
\catcode`@=11
\renewcommand{\section}{\@startsection{section}{1}{0pt}{\medskipamount}
{\medskipamount}{\Large\bf}}
\numberwithin{equation}{section}
\catcode`@=12

\begin{document}
\thispagestyle{empty}
\begin{flushright}
%\today\\
%Complete draft, version 3
\end{flushright}\vspace{1cm}
\begin{center}
{\LARGE\bf New $\cN{=}\,2$ superspace Calogero models}
\end{center}
\vspace{1cm}

\begin{center}
{\large\bf  Sergey Krivonos${}^{a}$, Olaf Lechtenfeld$^b$ and Anton Sutulin${}^a$}
\end{center}

\vspace{0.2cm}

\begin{center}
{${}^a$ \it
Bogoliubov  Laboratory of Theoretical Physics, JINR,
141980 Dubna, Russia}

${}^b$ {\it
Institut f\"ur Theoretische Physik and Riemann Center for Geometry and Physics, \\
Leibniz Universit\"at Hannover,
Appelstrasse 2, 30167 Hannover, Germany}

\vspace{0.5cm}

{\tt krivonos@theor.jinr.ru, lechtenf@itp.uni-hannover.de, sutulin@theor.jinr.ru}
\end{center}
\vspace{2cm}

\begin{abstract}\noindent
Starting from the Hamiltonian formulation of $\cN{=}\,2$ supersymmetric Calogero models
associated with the classical $A_n, B_n, C_n$ and $D_n$ series and their hyperbolic/trigonometric
cousins, we provide their superspace description. The key ingredients include $n$ bosonic and
$2n(n{-}1)$ fermionic $\cN{=}\,2$ superfields, the latter being subject to a nonlinear chirality constraint.
This constraint has a universal form valid for all Calogero models. With its help we find
more general supercharges (and a superspace Lagrangian), which provide the $\cN{=}\,2$ supersymmetrization
for bosonic potentials with arbitrary repulsive two-body interactions.
\end{abstract}


\vskip 0.5cm


\newpage
\setcounter{page}1
\section{Introduction}\noindent
The rational Calogero model \cite{Calogero, Calogero1}  and its various generalizations (see~\cite{Poly1} for a review) play a special role
in mathematical physics. Being an integrable and solvable multi-particle one-dimensional system it appears in many  different areas such as
fluid mechanics, spin chains, gauge theory and string theory. The standard $\cN{=}\,2$ supersymmetric extension of the $n$-particle  Calogero models is based on  $n$ scalar bosonic $\cN{=}\,2$ superfields with a component content $(1,2,1)$  i.e. one physical boson, 2 fermions and one auxiliary field for
each particle \cite{FM,DLM1,DLM2}. Being quite satisfactory for $\cN{=}\,2$ supersymmetry, this construction fails to reproduce $\cN{=}\,4$ supersymmetric Calogero models beyond four particles \cite{KL}.

Recent progress in the construction of supersymmetric extensions of Calogero models
was achieved by adding to the system more fermions as compared to the standard supersymmetrization~\cite{FM,DLM1,DLM2}.
It was inspired by supersymmetric extensions of the matrix models which, upon reduction or gauge fixing, give rise to the familiar bosonic systems.
The superfield approach developed in \cite{SF1,SF2,SF3} for the rational spin-Calogero models with $\cN{=}\,2,4$ supersymmetry was recently extended to
$\cN{=}\,2,4$ supersymmetric hyperbolic Calogero models~\cite{SF4,SF5}. However, it suffers from an unclear structure of the bosonic matrix model
one has to start from.

In the series of papers \cite{KLS1,KLPS,KL01} we developed a different approach. Mainly working in the Hamiltonian formulation,
we worked out an ansatz for the supercharges which accommodates all Calogero models associated with the classical
$A_n$, $B_n$, $C_n$ and $D_n$ Lie algebras and their trigonometric/hyperbolic extensions~\cite{KL01}.
Having at hands the Hamiltonian description of $\cN$-extended supersymmetric Calogero models, it is of interest to gain also its superfield formulation,
at least in the simplest case of $\cN{=}\,2$ supersymmetry.\footnote{
A superspace description of the $\cN$-extended supersymmetric Euler-Calogero-Moser system has been provided in \cite{KLS2}.}
The superspace picture, alternative to the one constructed in \cite{SF1,SF2,SF3,SF4,SF5}, will help in understanding the general supersymmetry structure and clarify the role played by the additional matrix fermions. This is the main goal of this work.

The plan of the paper is as follows. In Section~2 we review the Hamiltonian description of the supersymmetric Calogero models.
Their superfield treatment is performed in Section~3 ($A_1\oplus A_{n-1}$ models) and in Section~4 ($B_n$, $C_n$, $D_n$ models).
The most remarkable result here is a universal nonlinear fermionic chiral supermultiplet which collects all matrix fermions occurring in all
super-extended Calogero models.
In Section~5 we present more general supercharges (and a superspace Lagrangian), which provide an $\cN{=}\,2$ supersymmetrization
for a bosonic potential $\frac{g^2}{2} \sum_{i<j} f(x_i{-}x_j)^2$  with an arbitrary function $f$.
We conclude with a short summary and some arguments in favor of the integrability of the constructed supersymmetric systems.

\bigskip

\section{Hamiltonian description of $\cN{=}\,2$ supersymmetric Calogero models}\noindent
In the Hamiltonian approach the $n$-particle supersymmetric Calogero model
with $\cN{=}\,2$-extended supersymmetry~\cite{KL01, KLS1, KLPS} features the following degrees of freedom:
\begin{itemize}
\item  $n$ bosonic coordinates $x_i$ and momenta $p_i$ with $i=1,\ldots,n$,
\item  $ 2 \, n $ fermions $\psi_i$ and $\bpsi_{i}$ ,
\item $ 2\,n (n{-}1)$ fermions $\xi_{ij}$ and $\bxi_{ij}$ with $\xi_{ii}= \bxi_{ii}=0$.
\end{itemize}
Their non-vanishing Poisson brackets are
\be\label{PB1}
\left\{ x_i, p_j\right\}= \delta_{ij}, \qquad \left\{ \psi_i, \bpsi_{j} \right\} = -\im\, \delta_{ij}, \qquad
\left\{ \xi_{ij}, \bxi_{km}\right\} = -\im \left(1{-}\delta_{ij}\right)\left(1{-}\delta_{km}\right) \delta_{im}\delta_{jk}.
\ee
A central role of our construction take the composite objects $\Pi_{ij}$ and $\tPi_{ij}$ defined as
\bea
\Pi_{ij} \!\!&=&\!\! \left( \psi_i{-}\psi_j\right) \bxi_{ij}+\left( \bpsi_{i}{-}\bpsi_{j}\right) \xi_{ij}+
\sum_{k=1}^n \left( \xi_{ik}\bxi_{kj}+\bxi_{ik}\xi_{kj}\right), \label{pi} \\
\tPi_{ij} \!\!&=&\!\! 2 \delta_{ij} \psi_i \bpsi_i+\left( \psi_i{+}\psi_j\right) \bxi_{ij}-\left( \bpsi_{i}{+}\bpsi_{j}\right) \xi_{ij}+
\sum_{k=1}^n \left( \xi_{ik}\bxi_{kj}-\bxi_{ik}\xi_{kj}\right).
\eea
One may easily check that $\Pi_{ij}$ and $\tPi_{ij}$ together form an $s\!\left( u(n)\oplus u(n)\right)$ algebra,\footnote{
We remind that $\sum_i \Pi_{ii} =0$.}
\be\label{susus}
\begin{aligned}
&\big\{ \Pi_{ij}, \Pi_{km} \big\}=\im \big( \delta_{im} \Pi_{kj}-\delta_{kj}\Pi_{im}\big),\qquad
\big\{ \Pi_{ij}, \tPi_{km} \big\}=\im \big( \delta_{im} \tPi_{kj}-\delta_{kj}\tPi_{im}\big),\\
&\big\{ \tPi_{ij}, \tPi_{km} \big\}=\im \big( \delta_{im} \Pi_{kj}-\delta_{kj}\Pi\emph{}_{im}\big)\ .
\end{aligned}
\ee

Our $\cN{=}\,2$ supersymmetric Calogero models of $A$-type \cite{KL01,KLPS} are defined by a generic form of their supercharges,
\be\label{genQ}
\begin{aligned}
Q &\= \sum_{i=1}^n p_i \psi_{i} -\im\, \sum_{i\neq j}^n \Big[ \left( g + \Pi_{jj}\right)  f(z_{ij})+ \frac{ f{}'(z_{ij})}{f(z_{ij})}\Pi_{ij}\Big] \xi_{ji},\\
\bQ &\= \sum_{i=1}^n p_i \bpsi_{i} -\im\, \sum_{i\neq j}^n \Big[ \left( g + \Pi_{jj}\right)  f(z_{ij})+ \frac{ f{}'(z_{ij})}{f(z_{ij})}\Pi_{ij}\Big] \bxi_{ji},
\end{aligned}
\ee
with some function $f$, to be specified in a moment.
Note that $\tPi_{ij}$ does not appear here.
These supercharges form an $\cN{=}\,2$ super-Poincar\'{e} algebra,
\be\label{N2SP}
\big\{ Q , \bQ \big\} = - 2 \im\, \cH \und \big\{ Q, Q \big\}=\big\{ \bQ, \bQ \big\}=0 \ ,
\ee
together with the Hamiltonian
\be\label{H1}
\begin{aligned}
\cH \= \frac{1}{2}\sum_{i=1}^n p_i^2\ +\ \frac{1}{2}\sum_{i \neq j}^n \Big[ \left(g+\Pi_{jj}\right) f(z_{ij}) +\frac{f'(z_{ij})}{f(z_{ij})}\Pi_{ij}\Big]
\Big[ \left(g+\Pi_{ii}\right) f(z_{ij}) +\frac{f'(z_{ij})}{f(z_{ij})}\Pi_{ji}\Big]+ \frac{\alpha}{2} \sum_{i,j}^n \Pi_{ij}\Pi_{ji}.
\end{aligned}
\ee
Here, we abbreviated
\be
z_{ij}=x_i -x_j,
\ee
and the constant parameter $\alpha$ and the function $f$ are given as follows,
\be\label{f}
\begin{aligned}
\mbox{rational Calogero model} \qquad &  \alpha=0, \quad f(z_{ij}) =\frac{1}{z_{ij}}=\frac{1}{x_i{-}x_j}, \\
\mbox{hyperbolic  Calogero-Moser model} \qquad &  \alpha=-1, \quad f(z_{ij}) =\frac{1}{\sinh(z_{ij})}=\frac{1}{\sinh(x_i{-}x_j)}, \\
\mbox{trigonometric  Calogero-Moser model} \qquad &  \alpha=1, \quad f(z_{ij}) =\frac{1}{\sin(z_{ij})}=\frac{1}{\sin(x_i{-}x_j)}.
\end{aligned}
\ee
For the $B$, $C$ and $D$-type models, the supercharges take a more complicated generic form (including $\tPi_{ij}$),
\be\label{BCDQ}
\begin{aligned}
\cQ &\= \sum_{i=1}^n p_i \psi_{i} -\im\, \sum_{i\neq j}^n \Big[ \left( g + \Pi_{jj}\right)  f(z_{ij})+ \frac{ f{}'(z_{ij})}{f(z_{ij})}\Pi_{ij}\Big] \xi_{ji}+\im
\sum_{i\neq j}^n \Big[ \left( g + \Pi_{jj}\right)  f(y_{ij}) - \frac{ f{}'(y_{ij})}{f(y_{ij})}\tPi_{ij}\Big] \xi_{ji} \\
&\qquad + \im \sum_i^n \Big[ \left( g' + \Pi_{ii}\right)  f(y_{ii}) -\frac{ f{}'(y_{ii})}{f(y_{ii})}\tPi_{ii}\Big] \psi_{i}, \\
\cbQ &\= \sum_{i=1}^n p_i \bpsi_{i}  -\im\, \sum_{i\neq j}^n \Big[ \left( g + \Pi_{jj}\right)  f(z_{ij})+ \frac{ f{}'(z_{ij})}{f(z_{ij})}\Pi_{ij}\Big] \bxi_{ji}
-\im \sum_{i\neq j}^n\Big[ \left( g + \Pi_{jj}\right)  f(y_{ij}) - \frac{ f{}'(y_{ij})}{f(y_{ij})}\tPi_{ij}\Big] \bxi_{ji} \\
&\qquad - \im \sum_i^n \left[ \left( g' + \Pi_{ii}\right)  f(y_{ii}) -\frac{ f{}'(y_{ii})}{f(y_{ii})}\tPi_{ii}\right] \bpsi_{i}.
\end{aligned}
\ee
Here,
\be
y_{ij} = x_i+x_j ,
\ee
and the function $f$ is the same as in \p{f}. The supercharges \p{BCDQ} form the same $\cN{=}\,2$ super-Poincar\'{e} algebra \p{N2SP}
together with the Hamiltonian
\be\label{BCDH}
\begin{aligned}
\cH &\= \ \frac{1}{2}\sum_{i=1}^n p_i^2\ +\ \frac{1}{2}\sum_{i \neq j}^n \Big[ \left(g+\Pi_{jj}\right) f(z_{ij}) +\frac{f'(z_{ij})}{f(z_{ij})}\Pi_{ij}\Big]
\Big[ \left(g+\Pi_{ii}\right) f(z_{ij}) +\frac{f'(z_{ij})}{f(z_{ij})}\Pi_{ji}\Big]\\
&\qquad +  \frac{1}{2}\sum_{i \neq j}^n \Big[ \left(g+\Pi_{jj}\right) f(y_{ij}) -\frac{f'(y_{ij})}{f(y_{ij})}\tPi_{ij}\Big]
\Big[ \left(g+\Pi_{ii}\right) f(y_{ij}) -\frac{f'(y_{ij})}{f(y_{ij})}\tPi_{ji}\Big] \\
&\qquad + \frac{1}{2}\sum_{i}^n \Big[ \left(g'+\Pi_{ii}\right) f(y_{ii}) -\frac{f'(y_{ii})}{f(y_{ii})}\tPi_{ii}\Big]
\Big[ \left(g'+\Pi_{ii}\right) f(y_{ii}) -\frac{f'(y_{ii})}{f(y_{ii})}\tPi_{ii}\Big].
\end{aligned}
\ee
Its bosonic sector reads
\be\label{bosBCDH}
\cH_{bos} \= \frac{1}{2}\sum_{i=1}^n p_i^2\ +\frac{g^2}{2} \sum_{i\neq j}^n \left( f^2(z_{ij})+f^2(y_{ij})\right)+
\frac{g'{}^2}{2}\sum_{i}^n f^2(y_{ii}).
\ee
Due to the presence of only two coupling constants, $g$ and $g'$, we may describe $B$, $C$ and $D$-type models in the rational case and
$C$ and $D$ (but not $B$)-type models in the hyperbolic/trigonometric case.

\bigskip

\section{$\cN{=}\,2$ superspace $A_1 \oplus A_{n-1}$ Calogero models}\noindent
To provide a superspace description of $\cN{=}\,2$ supersymmetric Calogero models one has, firstly, \
to assemble the physical components $x_i$, $\psi_i$, $\bpsi_i$, $\xi_{ij}$ and $\bxi_{ij}$ into $\cN=2$ superfields.

It immediately follows  from the structure of the supercharges $Q$ and $\bQ$ \p{genQ} that under $\cN{=}\,2$ supersymmetry
the coordinates $x_i$ transform into the fermions $\psi_i$ and $\bpsi_i$:
\be
\delta x_i \equiv \left\{ x_i, \im\, \bar\epsilon Q + \im\, \epsilon \bQ\right\} =
\im\, \bar\epsilon \psi_i + \im\, \epsilon \bpsi_i\;.
\ee
Thus, one is let to $n$ bosonic $\cN{=}\,2$ superfields $\boldsymbol{x}_i$ with the components~\footnote{
We use the $\cN{=}\,2$ spinor covariant derivatives $D$ and $\bD$ obeying
$\left\{ D, \bD\right\} = 2 \im \partial_t$ and $\left\{D, D\right\} = \left\{\bD, \bD\right\}= 0.$
We denote by ${\cal A}|$ the $\theta=\bar{\theta} = 0$ limit of a superspace expression ${\cal A}$.
}
\be\label{compx}
x_i = \boldsymbol{x}_i|,\quad \psi_i =- \im D \boldsymbol{x}_i|,\quad \bpsi_i = -\im \bD \boldsymbol{x}_i|, \quad A_i = \sfrac{1}{2} \left[ \bD,D\right] \boldsymbol{x}_i|.
\ee

Concerning the fermionic components $\xi_{ij}, \bxi_{ij}$, we have no other possibility than to put them into
$2n(n{-}1)$ new fermionic superfields $\boldsymbol{\xi}_{ij}$ and $\bar{\boldsymbol{\xi}}_{ij}$ with vanishing diagonal parts, i.e.
\be
{\boldsymbol{\xi}}_{ii}=\bar{\boldsymbol{\xi}}_{ii}=0 \quad \forall\, i.
\ee
As $\cN{=}\,2$ superfields the $\boldsymbol{\xi}_{ij}$ and $\bar{\boldsymbol{\xi}}_{ij}$ contain a lot of components.
Hence, they have to be constrained somehow.\footnote{
It was first mentioned in~\cite{FIL-rev}  that $2n(n{-}1)$ fermionic components can be placed into
$n(n{-}1)$ supermultiplets of type $(0,2,2)$.}
The appropriate constraints derive from the explicit form
of the supercharges $Q$ and $\bQ$ \p{genQ}, which leads to the following supersymmetry transformations of the leading components \
$\xi_{ij}$ and $\bxi_{ij}$ of these superfields,
\be\label{tr1}
\begin{aligned}
\delta_Q  \xi_{ij} &\sim  \im \bar\epsilon\, \Big[ -\frac{f'(z_{ij})}{f(z_{ij})} \left(\psi_i{-}\psi_j\right)\, \xi_{ij}
+ \xi_{ij}\,\Big( \sum_{k \neq i}^n  f(z_{ik}) \xi_{ik}  -  \sum_{k \neq j}^n f(z_{jk}) \xi_{jk} \Big)
- \sum_{k \neq i,j}^n \Big(\, \frac{f'(z_{ik})}{f(z_{ik})}+\frac{f'(z_{kj})}{f(z_{kj})}\Big)\,\xi_{ik} \xi_{kj}
\Big]\,, \\
\delta_\bQ  \bxi_{ij} &\sim  \im \epsilon\, \Big[ -\frac{f'(z_{ij})}{f(z_{ij})} \left(\bpsi_i{-}\bpsi_j\right)\, \bxi_{ij}
+ \bxi_{ij}\,\Big( \sum_{k \neq i}^n  f(z_{ik}) \bxi_{ik}  -  \sum_{k \neq j}^n f(z_{jk}) \bxi_{jk} \Big)
- \sum_{k \neq i,j}^n \Big(\, \frac{f'(z_{ik})}{f(z_{ik})}+\frac{f'(z_{kj})}{f(z_{kj})}\Big)\,\bxi_{ik} \bxi_{kj}
\Big]\,.
\end{aligned}
\ee
To realize this transformation property we are forced to impose a nonlinear chirality condition on the superfields
$\boldsymbol{\xi}_{ij}$ and $\bar{\boldsymbol{\xi}}_{ij}$,
\be\label{chirality1}
\begin{aligned}
D \mxi_{ij} &= \im \, \Big[ -\frac{f'(\bb{z}_{ij})}{f(\bb{z}_{ij})} \left(\mpsi_i{-}\mpsi_j\right)\, \mxi_{ij}
+ \mxi_{ij}\,\Big( \sum_{k \neq i}^n  f(\bb{z}_{ik}) \mxi_{ik}  -  \sum_{k \neq j}^n f(\bb{z}_{jk}) \mxi_{jk} \Big)
- \sum_{k \neq i,j}^n \left(\, \frac{f'(\bb{z}_{ik})}{f(\bb{z}_{ik})}+\frac{f'(\bb{z}_{kj})}{f(\bb{z}_{kj})}\right)\,\mxi_{ik} \mxi_{kj}
\Big]\,, \\
\bD  \mbxi_{ij} &= \im \, \Big[ -\frac{f'(\bb{z}_{ij})}{f(\bb{z}_{ij})} \left(\mbpsi_i{-}\mbpsi_j\right)\, \mbxi_{ij}
+ \mbxi_{ij}\,\Big( \sum_{k \neq i}^n  f(\bb{z}_{ik}) \mbxi_{ik}  -  \sum_{k \neq j}^n f(\bb{z}_{jk}) \mbxi_{jk} \Big)
- \sum_{k \neq i,j}^n \Big(\, \frac{f'(\bb{z}_{ik})}{f(\bb{z}_{ik})}+\frac{f'(\bb{z}_{kj})}{f(\bb{z}_{kj})}\Big)\,\mbxi_{ik} \mbxi_{kj}
\Big]\,.
\end{aligned}
\ee
This condition leaves in the superfields $\boldsymbol{\xi}_{ij}$ and $\bar{\boldsymbol{\xi}}_{ij}$ only the components
\be\label{comprho}
\xi_{ij} = \bb{\xi}_{ij}|, \quad B_{ij} =\bD \bb{\xi}_{ij}|, \qquad \bxi_{ij} = \bb{\bxi}_{ij}|, \quad \bB_{ij} =D \bb{\bxi}_{ij}|\;.
\ee
Finally, to obtain the correct brackets \p{PB1} for $(\psi_i, \bpsi_i)$ and $(\xi_{ij}, \bxi_{ij})$ after passing to the Hamiltonian formalism,
the kinetic terms for these fermionic components must have the form
\be\label{kin1}
\cL^{\psi}_{kin} = \frac{\im}{2} \sum_{i=1}^n \Big( \dot{\psi}_i \bpsi_i - \psi_i \dot{\bpsi}_i \Big) \qquad\textrm{and}\qquad
\cL^{\xi}_{kin} = \frac{\im}{2} \sum_{i,j}^n \Big( \dot{\xi}_{ij} \bxi_{ji} - \xi_{ij} \dot{\bxi}_{ji} \Big).
\ee
In $\cN{=}\,2$ superspace, this amounts to the free action $(g=0)$
\be\label{action1}
S_0 \= \int\! dt\,d^2 \theta\ \Big[ -\frac{1}{2} \sum_{i=1}^n D \bb{x}_i \; \bD \bb{x}_i + \frac{1}{2} \sum_{i,j}^n \bb{\xi}_{ij} \bb{\bxi}_{ji} \Big]
\qquad\textrm{with}\quad d^2\theta \equiv D \bD.
\ee

More interesting is the construction of the interaction terms. Again, some hints come from the transformation properties of
the fermions $\xi_{ij}$ and $\bxi_{ij}$ under $\bQ$ and $Q$ supersymmetry, respectively,
\be\label{tr2}
\delta_\bQ  \xi_{ij} \sim  \im \epsilon\, g\, f(z_{ij})+\ldots, \quad  \delta_Q  \bxi_{ij} \sim  \im \bar\epsilon\, g\, f(z_{ij})+\ldots\, .
\ee
To reproduce such terms in superspoace, the unique possibility is  to add to the action $S_0$ \p{action1} a term
\be\label{action2}
S_{int} \= \im\,\frac{g}{2} \int\!dt\,d\bar\theta\ \sum_{i\neq j}^n f(\bb{z}_{ij})\bb{\xi}_{ij}
+\im\,\frac{g}{2} \int\!dt\,d\theta\ \sum_{i\neq j}^n f(\bb{z}_{ij})\,\bb{\bxi}_{ij}.
\ee
To be supersymmetrically invariant, the  integrands in \p{action2} must be chiral and antichiral, respectively.
It is not too hard to check that this is indeed so: the nonlinear chirality constraint \p{chirality1} implies that
\be\label{chir2}
D \Big(  \sum_{i\neq j}^n f(\bb{z}_{ij})\, \bb{\xi}_{ij}\Big) = 0 \qquad \mbox{ and } \qquad
\bD \Big( \sum_{i\neq j}^n f(\bb{z}_{ij})\,\bb{\bxi}_{ij}\Big) = 0.
\ee
Combining all these facts together, we conclude that the superfield action reads
\be\label{actionF}
S\= \int\!dt\,d^2 \theta\ \Big[ -\frac{1}{2} \sum_{i=1}^n D \bb{x}_i \; \bD \bb{x}_i + \frac{1}{2} \sum_{i,j}^n \bb{\xi}_{ij} \bb{\bxi}_{ji} \Big]
+\im\,\frac{g}{2} \int\!dt\,d\bar\theta\ \sum_{i\neq j}^n f(\bb{z}_{ij})\bb{\xi}_{ij}+\im\,\frac{g}{2} \int dt d\theta \sum_{i\neq j}^n f(\bb{z}_{ij})\,\bb{\bxi}_{ij}\,,
\ee
where the superfields $\bb{\xi}_{ij}$ and $\bb{\bxi}_{ij}$ are subject to the nonlinear chirality constraint \p{chirality1}.

It is important to note that, after passing to new fermionic superfields
\be\label{lambda1}
\bb{\lambda}_{ij} \equiv f(\bb{z}_{ij})\,\bb{\xi}_{ij} \qquad\textrm{and}\qquad
\bb{\blambda}_{ij} \equiv f(\bb{z}_{ij})\,\bb{\bxi}_{ij}\,,
\ee
the nonlinear constraint \p{chirality1} is slightly simplified to
\be\label{chirality2}
\begin{aligned}
D \boldsymbol{\lambda}_{ij}
&\= \im\,\Big[\,\bb{\lambda}_{ij} \sum_{k \neq i}^n  \bb{\lambda}_{ik} - \bb{\lambda}_{ij} \sum_{k \neq j}^n \bb{\lambda}_{jk}
+ \big (1{-}\delta_{ij} \big) \sum_{k \neq i,j}^n  \bb{\lambda}_{ik} \bb{\lambda}_{kj}\, \Big]\,, \\
\bD \boldsymbol{\blambda}_{ij}
&\= \im\,\Big[\, \bb{\blambda}_{ij} \sum_{k \neq i}^n \bb{\blambda}_{ik} - \bb{\blambda}_{ij} \sum_{k \neq j}^n \bb{\blambda}_{jk}
+ \big (1{-}\delta_{ij} \big) \sum_{k \neq i,j}^n  \bb{\blambda}_{ik} \bb{\blambda}_{kj}\, \Big]\,.
\end{aligned}
\ee
In this form, the constraint has lost any $f$-dependence, which however will reappear in the action,
\be\label{actionL}
S\= \int\!dt\,d^2 \theta\ \Big[ -\frac{1}{2} \sum_{i=1}^n D \bb{x}_i \; \bD \bb{x}_i
+ \frac{1}{2} \sum_{i,j}^n \frac{\bb{\lambda}_{ij} \bb{\blambda}_{ji}}{f(\bb{z}_{ij})\, f(\bb{z}_{ji})} \Big]
+\im\,\frac{g}{2} \int\!dt\,d\bar\theta\ \sum_{i\neq j}^n \bb{\lambda}_{ij}+\im\,\frac{g}{2} \int\!dt\,d\theta\ \sum_{i\neq j}^n \bb{\blambda}_{ij}\,.
\ee
Also, the component Lagrangian, Hamiltonian and Poisson brackets will be more complicated in terms of the composite superfields $\bb{\lambda}_{ij}$ and $\bb{\blambda}_{ij}$ .

Despite the extremely simple form of the superfield action \p{actionF}, its component version looks quite complicated
due to the constraint \p{chirality1}. Indeed, after integration over $\theta$ in \p{actionF} we get the off-shell Lagrangian
\be\label{cL1}
\cL = \cL_{0}+ \cL_{\textrm{pot}}\,, \qquad\textrm{where}
\ee
\be\label{cL2}
\begin{aligned}
\cL_{0} &\= \frac{1}{2}\, \sum_{i}^n \Big(\, {\dot x}_i {\dot x}_i + A_i A_i\, \Big) + \frac{\im}{2}\, \sum_{i}^n \Big( \dot{\psi}_i \bpsi_i - \psi_i \dot{\bpsi}_i \Big)
+ \frac{\im}{2}\, \sum_{i,j}^n \Big( \dot{\xi}_{ij} \bxi_{ji} - \xi_{ij} \dot{\bxi}_{ji} \Big)  \\
&\qquad + \frac{1}{2}\, \sum_{i,j}^n \Big(\, \xi_{ij} D \big( \bD \bxi_{ji} \big) - \bD \big( D \xi_{ij} \big) \bxi_{ji}
- D \xi_{ij} \bD \bxi_{ji} + B_{ij} \bB_{ji}\, \Big), \\
\cL_{\textrm{pot}} &\=- \frac{g}{2}\, \sum_{i,j}^n\, f'(z_{ij}) \Big(\,\left(\psi_i-\psi_j\right)\, \bxi_{ij} + \left(\bpsi_i-\bpsi_j\right)\, \xi_{ij} \Big)
+ \im\, \frac{g}{2}\, \sum_{i,j}^n \, f(z_{ij})\big(\, B_{ij} + \bB_{ij} \,\big) \,.
\end{aligned}
\ee
To eliminate the auxiliary fields $A_i$ and $B_{ij}$ one firstly has to evaluate the terms in the second line of \p{cL2} by using the
constraint \p{chirality1}. This is a straightforward but rather tedious calculation. After employing the  equations of motion for the auxiliary components we finally obtain the desired result,
\be\label{LN2fin}
\begin{aligned}
\cL &\= \frac{1}{2} \sum_{i=1}^n {\dot x}_i {\dot x}_i+
\frac{\im}{2} \sum_{i=1}^n \Big( \dot{\psi}_i \bpsi_i - \psi_i \dot{\bpsi}_i \Big)+ \frac{\im}{2} \sum_{i,j}^n \Big( \dot{\xi}_{ij} \bxi_{ji} - \xi_{ij} \dot{\bxi}_{ji}\Big) \\
&\qquad - \ \frac{1}{2}\sum_{i \neq j}^n \Big[ \big(g+\Pi_{jj}\big) f(z_{ij}) +\frac{f'(z_{ij})}{f(z_{ij})}\Pi_{ij}\Big]
\Big[ \big(g+\Pi_{ii}\big) f(z_{ij}) +\frac{f'(z_{ij})}{f(z_{ij})}\Pi_{ji}\Big]- \frac{\alpha}{2} \sum_{i,j}^n \Pi_{ij}\Pi_{ji}.
\end{aligned}
\ee
Thus, the superfield action \p{actionF} with the superfields $\bb{\xi}_{ij}$ and $ \bb{\bxi}_{ij}$  subject to the nonlinear chirality constraint \p{chirality1}
indeed describes all $\cN{=}\,2$ supersymmetric $A_1 \oplus A_{n-1}$ Calogero models.

\bigskip

\section{$\cN{=}\,2$ superspace $B_n$, $C_n$ and $D_n$ Calogero models}\noindent
The supercharges of the $\cN{=}\,2$ supersymmetric $B$, $C$ and $D$-type Calogero models \p{BCDQ} have a more complicated structure than those in~\p{genQ}.
Therefore, it is expected that the nonlinear chirality constraint for the superfields $\bb{\xi}_{ij}$ and $\bb{\bxi}_{ij}$ are more intricate as well.
Indeed, the explicit structure of the supercharges \p{BCDQ} uniquely fixes this constraint to be
\be\label{chiralityBCD}
\begin{aligned}
D \mxi_{ij} &\= \im \, \Big[-\frac{f'(\bb{z}_{ij})}{f(\bb{z}_{ij})} \big(\mpsi_i{-}\mpsi_j\big)\, \mxi_{ij}
-\frac{f'(\bb{y}_{ij})}{f(\bb{y}_{ij})} \big(\mpsi_i + \mpsi_j\big)\, \mxi_{ij} \\
&\qquad +\Big\{ \Big(\frac{f'(\bb{y}_{ii})}{f(\bb{y}_{ii})}-f(\bb{y_{ii}})\Big)\mpsi_i
+ \Big( \frac{f'(\bb{y}_{jj})}{f(\bb{y}_{jj})}+f(\bb{y}_{jj})\Big) \mpsi_j \Big\} \mxi_{ij} \\
&\qquad + \mxi_{ij}\,\Big( \sum_{k \neq i}^n \big( f(\bb{z}_{ik}) +f(\bb{y}_{ik}) \big)\mxi_{ik}
-  \sum_{k \neq j}^n \big(f(\bb{z}_{jk})+f(\bb{y}_{jk})\big)  \mxi_{jk} \Big) \\
&\qquad - \sum_{k \neq i,j}^n \Big(\, \frac{f'(\bb{z}_{ik})}{f(\bb{z}_{ik})}+\frac{f'(\bb{z}_{kj})}{f(\bb{z}_{kj})}-
\frac{f'(\bb{y}_{ik})}{f(\bb{y}_{ik})}+\frac{f'(\bb{y}_{kj})}{f(\bb{y}_{kj})}\Big)\,\mxi_{ik} \mxi_{kj}
\Big]\,, \\[8pt]
\bD \mbxi_{ij} &\= \im \, \Big[ -\frac{f'(\bb{z}_{ij})}{f(\bb{z}_{ij})} \big(\mbpsi_i{-}\mbpsi_j\big)\, \mbxi_{ij}
-\frac{f'(\bb{y}_{ij})}{f(\bb{y}_{ij})} \big(\mbpsi_i + \mbpsi_j\big)\, \mbxi_{ij} \\
&\qquad + \Big\{ \Big(\frac{f'(\bb{y}_{ii})}{f(\bb{y}_{ii})}+f(\bb{y_{ii}})\Big)\mbpsi_i
+ \Big( \frac{f'(\bb{y}_{jj})}{f(\bb{y}_{jj})}-f(\bb{y}_{jj})\Big) \mbpsi_j\Big\} \mbxi_{ij} \\
&\qquad + \mbxi_{ij}\,\Big( \sum_{k \neq i}^n \big( f(\bb{z}_{ik}) -f(\bb{y}_{ik}) \big)\mbxi_{ik}
- \sum_{k \neq j}^n \big( f(\bb{z}_{jk})-f(\bb{y}_{jk})\big)  \mbxi_{jk} \Big) \\
&\qquad - \sum_{k \neq i,j}^n \Big(\, \frac{f'(\bb{z}_{ik})}{f(\bb{z}_{ik})}+\frac{f'(\bb{z}_{kj})}{f(\bb{z}_{kj})}-
\frac{f'(\bb{y}_{ik})}{f(\bb{y}_{ik})}+\frac{f'(\bb{y}_{kj})}{f(\bb{y}_{kj})}\Big)\,\mbxi_{ik} \mbxi_{kj}
\Big]\,.
\end{aligned}
\ee
The complicated form of this constraint disappears after passing to the composite superfields
\be\label{lambdaBCD}
\bb{\lambda}_{ij} \=\big( f(\bb{z}_{ij})+f(\bb{y}_{ij})\big) \bb{\xi}_{ij} \qquad\textrm{and}\qquad
\bb{\blambda}_{ij} \=\big( f(\bb{z}_{ij})-f(\bb{y}_{ij})\big) \bb{\bxi}_{ij},
\ee
in which it acquires its familiar form \p{chirality2},
\be
\begin{aligned}
D \boldsymbol{\lambda}_{ij}
&\= \im\,\Big[\,\bb{\lambda}_{ij} \sum_{k \neq i}^n  \bb{\lambda}_{ik} - \bb{\lambda}_{ij} \sum_{k \neq j}^n \bb{\lambda}_{jk}
+ \big (1{-}\delta_{ij} \big) \sum_{k \neq i,j}^n  \bb{\lambda}_{ik} \bb{\lambda}_{kj}\, \Big]\,, \\
\bD \boldsymbol{\blambda}_{ij}
&\= \im\,\Big[\, \bb{\blambda}_{ij} \sum_{k \neq i}^n \bb{\blambda}_{ik} - \bb{\blambda}_{ij} \sum_{k \neq j}^n \bb{\blambda}_{jk}
+ \big (1{-}\delta_{ij} \big) \sum_{k \neq i,j}^n  \bb{\blambda}_{ik} \bb{\blambda}_{kj}\, \Big]\,. \nn
\end{aligned}
\ee
Finally, the superfield action reads
\be\label{actionBCD}
\begin{aligned}
S&\= \int\!dt\,d^2\theta\ \Big[ -\frac{1}{2} \sum_{i=1}^n D \bb{x}_i \; \bD \bb{x}_i
+ \frac{1}{2} \sum_{i,j}^n \bb{\xi}_{ij} \bb{\bxi}_{ji} - \frac{1}{2}\, g'\, h(\bb{y}_{ii})\Big] \\
&\qquad +\im\,\frac{g}{2} \int\!dt\,d\bar\theta\ \sum_{i\neq j}^n\big( f(\bb{z}_{ij})+f(\bb{y}_{ij})\big)\bb{\xi}_{ij}+
\im\,\frac{g}{2} \int\!dt\,d\theta\ \sum_{i\neq j}^n \big( f(\bb{z}_{ij})+f(\bb{y}_{ij})\big)\,\bb{\bxi}_{ij}\, ,
\end{aligned}
\ee
where
\be\label{h}
h'(\bb{y}_{ii}) = f(\bb{y}_{ii}).
\ee
Compared to the action of the $A_1 \oplus A_{n-1}$ Calogero models \p{actionL}, only the term
$\frac{1}{2}\,g'\int\!dt\,d^2\theta\ h(\bb{y}_{ii})$ carrying the new coupling constant $g'$ appears in the action \p{actionBCD}.
All other terms just mimic those in~\p{actionL}.

It is a matter of straightforward but tedious calculations to check that, after excluding the auxiliary fields by their equations of motion,
the final Lagrangian acquires the expected form
\be\label{actionBCDH}
\begin{aligned}
\cL &\= \frac{1}{2} \sum_{i=1}^n {\dot x}_i {\dot x}_i+
\frac{\im}{2} \sum_{i=1}^n \Big( \dot{\psi}_i \bpsi_i - \psi_i \dot{\bpsi}_i \Big)+ \frac{\im}{2} \sum_{i,j}^n \Big( \dot{\xi}_{ij} \bxi_{ji} - \xi_{ij} \dot{\bxi}_{ji}\Big) \\
&\qquad -\frac{1}{2}\sum_{i \neq j}^n \Big[ \big(g+\Pi_{jj}\big) f(z_{ij}) +\frac{f'(z_{ij})}{f(z_{ij})}\,\Pi_{ij}\Big]
\Big[ \big(g+\Pi_{ii}\big) f(z_{ij}) +\frac{f'(z_{ij})}{f(z_{ij})}\,\Pi_{ji}\Big]\\
&\qquad -\frac{1}{2}\sum_{i \neq j}^n \Big[ \big(g+\Pi_{jj}\big) f(y_{ij}) -\frac{f'(y_{ij})}{f(y_{ij})}\,\tPi_{ij}\Big]
\Big[ \big(g+\Pi_{ii}\big) f(y_{ij}) -\frac{f'(y_{ij})}{f(y_{ij})}\,\tPi_{ji}\Big] \\
&\qquad -\frac{1}{2}\sum_{i}^n \Big[ \big(g'+\Pi_{ii}\big) f(y_{ii}) -\frac{f'(y_{ii})}{f(y_{ii})}\,\tPi_{ii}\Big]
\Big[ \big(g'+\Pi_{ii}\big) f(y_{ii}) -\frac{f'(y_{ii})}{f(y_{ii})}\,\tPi_{ii}\Big].
\end{aligned}
\ee
Thus, the superfield action \p{actionBCD} with the superfields $\bb{\xi}_{ij}$ and $ \bb{\bxi}_{ij}$ subject to the nonlinear chirality constraint \p{chiralityBCD}
indeed describes the $\cN{=}\,2$ supersymmetric $B$, $C$ and $D$-type Calogero models.\footnote{
Except again for $B$-type models in the trigonometric/hyperbolic case.}

\bigskip

\section{New $\cN{=}\,2$ supersymmetric $n$-particle models}\noindent
Our point of departure was the explicit form of the $\cN{=}\,2$ supercharges for the Calogero models \p{genQ} and \p{BCDQ} constructed in \cite{KL01, KLPS}.
In all cases we considered the function $f$ to be not arbitrary but to be chosen from the list~\p{f}.
Indeed, for generic $f$ the supercharges \p{genQ} do not form the closed algebra~\p{N2SP}. As a consequence,
the nonlinear chirality condition~\p{chirality1} is not self-consistent for an arbitrary function $f$
(i.e.\ $D$ acting on the r.h.s.\ of~\p{chirality1} does not vanish). On the other hand, the universal chirality constraint~\p{chirality2} is self-consistent.
This implies a weaker chirality condition on $\bb{\xi}_{ij}$ and $\bb{\bxi}_{ij}$:\footnote{
For the sake of simplicity we consider here the $A$-type case,
in which the $(\bb{\lambda}_{ij},\bb{\blambda}_{ij})$ and $(\bb{\xi}_{ij},\bb{\bxi}_{ij})$ are related as in \p{lambda1}.}
\be\label{chirality3}
\begin{aligned}
D \mxi_{ij} &\= \im \, \Big[ -\frac{f'(\bb{z}_{ij})}{f(\bb{z}_{ij})} \left(\mpsi_i -\mpsi_j\right)\, \mxi_{ij}
+ \mxi_{ij}\,\Big( \sum_{k \neq i}^n  f(\bb{z}_{ik}) \mxi_{ik}  -  \sum_{k \neq j}^n f(\bb{z}_{jk}) \mxi_{jk} \Big)
+\sum_{k \neq i,j}^n \, \frac{f(\bb{z}_{ik})\,f(\bb{z}_{kj})}{f(\bb{z}_{ij})}\,\mxi_{ik} \mxi_{kj} \Big]\,, \\
\bD \mbxi_{ij} &\= \im \, \Big[ -\frac{f'(\bb{z}_{ij})}{f(\bb{z}_{ij})} \left(\mbpsi_i -\mbpsi_j\right)\, \mbxi_{ij}
+ \mbxi_{ij}\,\Big( \sum_{k \neq i}^n  f(\bb{z}_{ik}) \mbxi_{ik}  -  \sum_{k \neq j}^n f(\bb{z}_{jk}) \mbxi_{jk} \Big)
+\sum_{k \neq i,j}^n \, \frac{f(\bb{z}_{ik})\,f(\bb{z}_{kj})}{f(\bb{z}_{ij})}\,\mbxi_{ik} \mbxi_{kj}
\Big]\,.
\end{aligned}
\ee
One may check that this nonlinear chirality constraint is perfectly self-consistent for an {\it arbitrary\/} function $f$.

Now, let us again start from the superspace action \p{action1}, but where the fermionic superfields $\bb{\xi}_{ij}$ and $\bb{\bxi}_{ij}$ must obey
the constraint~\p{chirality3}. Passing to the components and eliminating the auxiliary components one arrives at
\be\label{genQQ}
\begin{aligned}
\cQ &\= \sum_{i=1}^n p_i \psi_{i} -\im\, \sum_{i\neq j}^n \Big[ \left( g + \Pi_{jj}\right)  f(z_{ij})+
\frac{ f{}'(z_{ij})}{f(z_{ij})}\left(\psi_i-\psi_j\right) \bxi_{ij} + \sum_{k\neq i,j}^n \frac{f(z_{ik})\, f(z_{ij})}{f(z_{kj})} \xi_{ik} \bxi_{kj} \Big] \xi_{ji}, \\
\cbQ &\= \sum_{i=1}^n p_i \bpsi_{i} -\im\, \sum_{i\neq j}^n \Big[ \left( g + \Pi_{jj}\right)  f(z_{ij})+
\frac{ f{}'(z_{ij})}{f(z_{ij})}\left(\bpsi_i-\bpsi_j\right) \xi_{ij} + \sum_{k\neq i,j}^n \frac{f(z_{ik})\, f(z_{ij})}{f(z_{kj})} \bxi_{ik} \xi_{kj} \Big] \bxi_{ji}.
\end{aligned}
\ee
For the functions listed in \p{f} these supercharges coincide with the ones in \p{genQ}. However, the supercharges \p{genQQ} generate the
$\cN{=}\,2$ super-Poincar\'{e} algebra \p{N2SP} for an {\it arbitrary\/} function~$f$. It is not too hard to find the bosonic part of the new Hamiltonian,
\be\label{genHH}
\cH_{\textrm{bos}} = \frac{1}{2} \sum_{i}^n p_i^2 + \frac{g^2}{2} \sum_{i\neq j}^n f^2(z_{ij})\,.
\ee
Thus, the supercharges \p{genQQ} provide an $\cN{=}\,2$ supersymmetrization of a wide class of multi-particle systems with a bosonic Hamiltonian of the
type~\p{genHH}. A detailed analysis of such models will be given elsewhere.

\bigskip

\section{Conclusions}\noindent
In this paper we have provided a superspace description of the $\cN{=}\,2$ supersymmetric Calogero models, rational as well as trigonometric/hyperbolic,
associated with the classical $A_n$, $B_n$, $C_n$ and $D_n$ Lie algebras. We presented a minimal superfield content accomodating the $2n^2$ fermions
for the $\cN{=}\,2$ supersymmetric $n$-particle model. As $2n$ fermions accompany the $n$ bosonic coordinates in general bosonic $\cN{=}\,2$ superfields,
the remaining $2n(n{-}1)$ fermions must be put into additional fermionic $\cN{=}\,2$ superfields, which have to be constrained such as to describe those fermions alone.
The nonlinear chirality condition~\p{chirality2} written in terms of commposite superfields solves this task.
These composite fermionic $\cN{=}\,2$ superfields make the constraint look simple and universal but complicate the Lagrangian.
In terms of the fundamental fermions it is the other way around.
We finally presented more general supercharges (and the superspace Lagrangian) which provides an $\cN{=}\,2$ supersymmetrization of
bosonic $n$-particle systems with an arbitrary repulsive two-body interaction.

One might criticize that the approach presented here is unnecessarily complicated, because all $\cN{=}\,2$ supersymmetric Calogero models
can be more or less straightforwardly formulated in the standard fashion employing the minimum of $2n$ fermions~\cite{FM}.
This, however, is no longer the case with $\cN{>}\,2$ supersymmetric Calogero models, where our treatment with additional fermions becomes essential.
Hence, we consider our results here as a preparation for attacking Calogero systems with more supersymmetry in a superspace setting.
The main excuse for presenting of our $\cN{=}\,2$ results is the universal form of the nonlinear chirality constraint together with the almost trivial generalization
of the supercharges to $\cN{=}\,4$ supersymmetry~\cite{KL01} which will further the superspace construction of $\cN{\ge}\,4$ Calogero models.

There is no doubt that the importance of Calogero-type models is related with their integrability. In the standard case when $\cN{=}\,2$ supersymmetrization of $n$-particle models involves $2 n$ fermionic degrees of freedom the issue of  integrability has been considered, for example, in \cite{DLM1,DLM2}. In our case
we deal with many more fermionic degrees of freedom, namely $2 n^2$ instead of $2 n$. Nevertheless, the $L{-}M$ pair, at least for rational Calogero models ($\alpha=0, f(z_{ij})=\frac{1}{x_i-x_j}$ in \p{f}), has the quite simple form
\bea\label{lax1}
&& L_{ij} = \delta_{ij}\, p_j -\im \left(1-\delta_{ij}\right) \frac{g+ \Pi_{jj}-\Pi_{ij}}{x_i-x_j}, \nn \\
&& M_{ij} = -\im \, \delta_{ij} \sum_{k \neq i} \frac{g+\Pi_{kk}-\Pi_{ik}}{(x_i-x_k)^2} +\im \left( 1- \delta_{ij}\right) \frac{g+\Pi_{jj}-\Pi_{ij}}{(x_i-x_j)^2}
\eea
such that
\be\label{laxeq1}
\frac{d}{dt} L_{ij} \equiv \left\{ L_{ij}, H \right\} = \sum_k \left( L_{ik} \, M_{kj} - M_{ik}\, L_{kj}\right) \equiv \left[ L, M\right]_{ij}.
\ee
This construction is quite similar to those  considered in  \cite{DLM1,DLM2} and \cite{SF5}.

Clearly, the independent conserved   currents read
\be\label{Ik}
{\cH}_k =  \mbox{Tr } \left( L^k\right), \quad k= 1,2,\ldots, n.
\ee
Additional currents can be constructed if we note that
\be\label{eqxi}
\frac{d}{dt} \Pi_{ij} \equiv  \left\{ \Pi_{ij}, H \right\} = \left[ \Pi ,\, M \right]_{ij} \;  .
\ee
Therefore, any operator that can be written as the trace of a polynomial function $\cF$ depending only on the matrices $L$ and $\Pi$ is conserved,
\be\label{curr1}
\frac{d}{dt} \tr \cF (L, \Pi) = 0.
\ee
Keeping in mind that our Hamiltonians \p{H1} depend on  fermionic degrees of freedom only through the composites $\Pi_{ij}$, one  expects that the above set of  conserved currents contains all currents in  involution, as needed for  integrability. This will be a strong indication in favor of complete  integrability of the constructed systems. However, a detailed discussion of integrability is out of the scope of the present paper.

\vspace{0.5cm}

\noindent{\large\bf Acknowledgements}\\
\noindent
We thank Armen Nersessian for many fruitful discussions.
This work was supported by the RFBR-DFG grant No 20-52-12003.
The work of A.S. was partially supported by RFBR grant  18-02-01046.

\bigskip

%\setcounter{equation}{0}
%\def\theequation{A.\arabic{equation}}
%\section*{Appendix A.}
\appendix
\section{Appendix}\noindent
The nonlinear chirality constraint reads
\bea
&& \sfrac{1}{\im}\,D \mlambda_{ij} \=  \mlambda_{ij}\, \sum_{k \neq i} \mlambda_{ik}  - \mlambda_{ij}\, \sum_{k \neq j} \mlambda_{jk}
+ \big(1{-}\delta_{ij} \big) \,\sum_{k \neq i,j} \mlambda_{ik} \mlambda_{kj}\,,\qquad  \mlambda_{ii} = 0\,, \label{A1}\\
&& \sfrac{1}{\im}\,\overline D \mblambda_{ij} \=  \mblambda_{ij}\, \sum_{k \neq i} \mblambda_{ik}  - \mblambda_{ij}\, \sum_{k \neq j} \mblambda_{jk}
+ \big(1{-}\delta_{ij} \big) \,\sum_{k \neq i,j} \mblambda_{ik} \mblambda_{kj}\,, \qquad  \mblambda_{ii} = 0\,.
\eea
Let us prove that \p{A1} is self-consistent.
For this we act with $D$ on it. The l.h.s.\ is identically zero (since $D^2=0$), while the r.h.s.\ should vanish
by virtue of the constraint itself.
\be\label{A2}
\begin{aligned}
\textrm{r.h.s.} &\= D \mlambda_{ij}\, \sum_{k \neq i} \mlambda_{ik}  - \mlambda_{ij}\, \sum_{k \neq i} D \mlambda_{ik}
- D \mlambda_{ij}\, \sum_{k \neq j} \mlambda_{jk}  + \mlambda_{ij}\,
\sum_{k \neq j} D \mlambda_{jk} %\\
+ \big(1{-}\delta_{ij} \big) \,\sum_{k \neq i,j} \big(D \mlambda_{ik} \mlambda_{kj} - \mlambda_{ik} D \mlambda_{kj} \big) \\
&\= \Big(
\mlambda_{ij}\, \sum_{n \neq i} \mlambda_{in}  - \mlambda_{ij}\, \sum_{n \neq j} \mlambda_{jn}
+ \big(1{-}\delta_{ij} \big) \,\sum_{n \neq i,j} \mlambda_{in} \mlambda_{nj}
\Big)
\Big[ \sum_{k \neq i} \mlambda_{ik} - \sum_{k \neq j} \mlambda_{jk} \Big] \\
&\qquad - \mlambda_{ij} \sum_{k \neq i} \Big(
\mlambda_{ik}\, \sum_{n \neq i} \mlambda_{in}  - \mlambda_{ik}\, \sum_{n \neq k} \mlambda_{kn}
+ \big(1{-}\delta_{ik} \big) \,\sum_{n \neq i,k} \mlambda_{in} \mlambda_{nk} \Big) \\
&\qquad +
\mlambda_{ij} \sum_{k \neq j} \Big(
\mlambda_{jk}\, \sum_{n \neq j} \mlambda_{jn}  - \mlambda_{jk}\, \sum_{n \neq k} \mlambda_{kn}
+ \big(1{-}\delta_{jk} \big) \,\sum_{n \neq j,k} \mlambda_{jn} \mlambda_{nk} \Big) \\
&\qquad +
\big(1{-}\delta_{ij} \big) \, \sum_{k \neq i,j} \Big(
\mlambda_{ik}\, \sum_{n \neq i} \mlambda_{in}  - \mlambda_{ik}\, \sum_{n \neq k} \mlambda_{kn}
+ \big(1{-}\delta_{ik} \big) \,\sum_{n \neq i,k} \mlambda_{in} \mlambda_{nk} \Big) \, \mlambda_{kj} \\
&\qquad -
\big(1{-}\delta_{ij} \big) \, \sum_{k \neq i,j} \Big(
\mlambda_{kj}\, \sum_{n \neq k} \mlambda_{kn}  - \mlambda_{kj}\, \sum_{n \neq j} \mlambda_{jn}
+ \big(1{-}\delta_{kj} \big) \,\sum_{n \neq j,k} \mlambda_{kn} \mlambda_{nj} \Big) \, \mlambda_{ik}\,.
\end{aligned}
\ee

Let us first collect all terms proportional to $(1{-}\delta_{ij})$,
\be\label{A3}
\begin{aligned}
&\sum_{n \neq i,j} \mlambda_{in} \mlambda_{nj} \sum_{k \neq i} \mlambda_{ik}
- \sum_{n \neq i,j} \mlambda_{in} \mlambda_{nj} \sum_{k \neq j}
\mlambda_{jk}
+ \sum_{k \neq i,j} \mlambda_{ik} \sum_{n \neq i} \mlambda_{in} \mlambda_{kj}
- \sum_{k \neq i,j} \mlambda_{ik} \sum_{n \neq k} \mlambda_{kn}
\mlambda_{kj} \\
+ &
\sum_{k \neq i,j} \sum_{n \neq i,k} \mlambda_{in}  \mlambda_{nk} \mlambda_{kj}
- \sum_{k \neq i,j} \mlambda_{kj} \sum_{n \neq k} \mlambda_{kn}
\mlambda_{ik}
+ \sum_{k \neq i,j} \mlambda_{kj} \sum_{n \neq j} \mlambda_{jn} \mlambda_{ik}
- \sum_{k \neq i,j} \sum_{n \neq k,j} \mlambda_{kn}
\mlambda_{nj} \mlambda_{ik}\,.
\end{aligned}
\ee
The fourth and sixth terms cancel each other out.
Then, replacing the indices in the first and second terms as $k\leftrightarrow n$ makes them cancel with the third and seventh terms.
Thus, after replacing the indices in the first term as $k\leftrightarrow n$, we remain with
\be\label{A5}
\sum_{n \neq i,j} \sum_{k \neq i,n} \mlambda_{ik}  \mlambda_{kn} \mlambda_{nj}
- \sum_{k \neq i,j} \sum_{n \neq k,j} \mlambda_{ik}
\mlambda_{kn}  \mlambda_{nj}\,.
\ee
A sum over $k$ in the first term can be split into two pieces, corresponding to $k=j$ or $k\neq j$, yielding
\be\label{A6}
\sum_{n \neq i,j} \sum_{k \neq i,n} \mlambda_{ik}  \mlambda_{kn} \mlambda_{nj} =
\sum_{n \neq i,j} \mlambda_{ij}  \mlambda_{jn} \mlambda_{nj}
+\sum_{n \neq i,j} \sum_{k \neq i,j,n} \mlambda_{ik}  \mlambda_{kn} \mlambda_{nj}\,.
\ee
Performing the analogous splitting for the index $n$ in the second term of \p{A5}, $n=i$ or $n\neq i$, one finds
\be\label{A7}
\sum_{k \neq i,j} \sum_{n \neq k,j} \mlambda_{ik}
\mlambda_{kn} \mlambda_{nj} =
\sum_{k \neq i,j} \mlambda_{ik}  \mlambda_{ki}
\mlambda_{ij}
+ \sum_{k \neq i,j} \sum_{n \neq i,j,k} \mlambda_{ik}
\mlambda_{kn} \mlambda_{nj}\,.
\ee
After relabeling $k\leftrightarrow n$, the second terms on the r.h.s.\ of \p{A6} and \p{A7} become identical.
Hence, their contribution to \p{A5} cancels,
\be\label{A8}
\sum_{k \neq i,j} \sum_{n \neq i,j,k} \Big (\mlambda_{in}  \mlambda_{nk} \mlambda_{kj}
- \mlambda_{ik}  \mlambda_{kn} \mlambda_{nj} \Big) =
0\,,
\ee
so we are left with the first terms on the r.h.s.\ of \p{A6} and \p{A7},
\be\label{A9}
\mlambda_{ij} \sum_{n \neq i,j} \Big (\mlambda_{jn} \mlambda_{nj} - \mlambda_{in}  \mlambda_{ni} \Big)\,.
\ee

Let us secondly analyze all terms in \p{A2} multiplying $\mlambda_{ij}$ and add to them \p{A9}, which is of the same form,
\be\label{A10}
\begin{aligned}
&
\sum_{n \neq i} \mlambda_{in} \sum_{k \neq i} \mlambda_{ik}
- \sum_{n \neq i} \mlambda_{in} \sum_{k \neq j} \mlambda_{jk}
- \sum_{n \neq j} \mlambda_{jn} \sum_{k \neq i} \mlambda_{ik}
+ \sum_{n \neq j} \mlambda_{jn} \sum_{k \neq j} \mlambda_{jk}
\\
-&\sum_{k \neq i} \mlambda_{ik} \sum_{n \neq i} \mlambda_{in}
+ \sum_{k \neq i} \mlambda_{ik} \sum_{n \neq k} \mlambda_{kn}
- \sum_{k \neq i} \sum_{n \neq i,k} \mlambda_{in} \mlambda_{nk}
\\
+&\sum_{k \neq j} \mlambda_{jk} \sum_{n \neq j} \mlambda_{jn}
- \sum_{k \neq j} \mlambda_{jk} \sum_{n \neq k} \mlambda_{kn}
+ \sum_{k \neq j} \sum_{n \neq j,k} \mlambda_{jn} \mlambda_{nk}
\\
+&\sum_{k \neq i,j}  \mlambda_{jk} \mlambda_{kj}
- \sum_{k \neq i,j} \mlambda_{ik}  \mlambda_{ki} \,.
\end{aligned}
\ee
It remains to show that these twelve terms combine to zero.
The sum of the first five term cancels with the eighth term.
Rewriting the sixth term as a sum of two pieces, separating $n=i$ and $n\neq i$,
\be\label{A13}
\sum_{k \neq i} \mlambda_{ik} \sum_{n \neq k} \mlambda_{kn}
\= \sum_{k \neq i} \mlambda_{ik} \mlambda_{ki} + \sum_{k \neq i} \mlambda_{ik} \sum_{n \neq k,i} \mlambda_{kn}\,,
\ee
one may observe it to cancel with the seventh and the twelveth term.
Analogously, the ninth, tenth and eleventh term add to zero.
Nothing remains. This completes the proof.

\end{document}